\documentclass[12pt]{aastex}
\usepackage{natbib}
\usepackage{graphicx}
\usepackage{epsf,amsfonts,amsmath,amssymb}
\usepackage{mathrsfs}
\usepackage[usenames,dvipsnames,svgnames,table]{xcolor}
\usepackage{ulem,color}
\usepackage{array}
\usepackage{epstopdf}
\usepackage{multirow}
\usepackage{ulem}
\usepackage{rotating}

\begin{document}

\author{Paz Beniamini, Kenta Hotokezaka and Tsvi Piran}
\affil{ Racah Institute of Physics, The Hebrew University of Jerusalem, Jerusalem 91904, Israel
}

\title{Natal Kicks and Time Delays in Merging Neutron Star Binaries -  Implications for $r$-process nucleosynthesis in Ultra Faint Dwarfs and in the Milky Way}

\begin{abstract}
Merging neutron star binaries are prime candidate sources for heavy $r$-process nucleosynthesis. The amount of heavy $r$-process material is consistent with the mass ejection and rates of mergers, and abundances of relic radioactive materials suggest that heavy $r$-process material is produced in rare events. Observations of possible macronovae provide further support for this model. Still, some concerns remain. One is the observation of heavy $r$-process elements in Ultra Faint Dwarf (UFD) galaxies. The escape velocities from UFDs are so small that the natal kicks, taking place at neutron stars' birth, might eject such binaries from UFDs. Furthermore the old stellar populations of UFDs requires that  $r$-process nucleosynthesis must have taken place very early on, while it may take several Gyr for compact binaries to merge. This last problem arises also within the Milky Way where heavy $r$-process materials has been observed in some low metallicity stars. We show here that {$\gtrsim 0.5$} of neutron star binaries form with a {sufficiently} small proper motion {to remain bound even in} a UFD. Furthermore, {approximately $90\%$ of DNSs with an initial separation of $10^{11}$cm merge within 300Myrs and $\approx 15\%$ merge in less than 100Myrs.}  
This population of ``rapid mergers"  explains the  appearance of heavy $r$-process material in both UFDs and in the early Milky Way. 
\end{abstract}

\section{Introduction}
\label{sec:Introduction}

Merging double neutron stars (DNSs) 
that eject highly neutron rich material are prime candidates for the production sites of heavy $r$(apid)-process elements \citep{lattimer1976ApJ, eichler1989Nature,freiburghaus1999ApJ}. 
The overall amount of heavy $r$-process material in the Milky Way is consistent with
the expectations of mass ejection in numerical merger simulations \cite[e.g.][]{sekiguchi2015PRD,radice2016}
with their expected rates as estimated from Galactic DNSs \cite[see e.g.][]{kim2015MNRAS}
or from the rate \citep{GuettaPiran06,wanderman2015MNRAS} of short Gamma-Ray Bursts (sGRBs).
Discoveries of $r$-process driven macronova~(kilonova) candidates associated with 
sGRBs ~\citep{tanvir2013Nature, berger2013ApJ,yang2015NatCo, jin2016} 
provided further observational evidence of the DNS merger scenario \cite[e.g.][]{piran2014}.
{Following these developments, several recent works \citep{shen2015ApJ, vandevoort2015MNRAS,wehmeyer2015MNRAS,Montes16}
have shown that under reasonable assumptions DNS mergers can account for the history of R-process enrichment in the Galaxy.}

However, recently, \cite{Bramante(2016)} have used the observations of  $r$-process elements in dwarf satellite galaxies to question
the DNS merger scenario for  $r$-process production.
\cite{ji2015}, and independently, \cite{roederer2016} reported  the discovery 
of an $r$-process enriched ultra-faint dwarf~(UFD) galaxy Reticulum II, the total stellar luminosity of Reticulum II
is $\sim 1000L_{\odot}$ and the {line of sight} velocity dispersion is $\sim 4$~km/s~\citep{walker2015ApJ}.
\cite{Bramante(2016)} suggested that the kick given to the DNS during the second collapse would eject the binary from such a small galaxy.
{A second problem that arises is that} UFDs are composed of very old stellar population~\citep{brown2014ApJ,weisz2015ApJ}, suggesting that 
the chemical abundances have been frozen since $\approx 13$~Gyr ago. This
requires that the $r$-process formation should take place relatively soon after the formation of the first stars.  
This raises the question whether mergers could take place sufficiently rapidly so that their $r$-process material would be able to enrich the old stellar population.
\cite{Bramante(2016)} suggested, therefore, that  
a different mechanism must have produced the observed $r$-process material in these galaxies.

A significant population of ``rapid mergers" ($<$Gyr) is natural, and in fact is expected from observations of DNS systems in our Galaxy. Two of the ten observed DNS systems in our Galaxy (that don't reside in globular clusters and for which the masses are well constrained), the double pulsar, J0737-3039, and the original binary pulsar, B1913-16, will merge in less than a few hundred Myr.
Given the spin down time of the pulsars in these systems, we can constrain the {life time of these systems since the formation of the DNS to less than 140\,Myr and 460\,Myr respectively.}
{Indeed the existence rapid mergers has previously been suggested using population synthesis models 
\citep{Belczynski(2002),Belczynski2007,O'Shaughnessy2008ApJ}}.
Furthermore, the observed small proper motion of J0737-3039,  $\lesssim 10 \mbox{~km s}^{-1}$ \citep{Kramer2006Sci}, implies that some ``rapid mergers" {move slowly} enough to remain confined even within UFDs \footnote{{Evolving J0737-3039 backwards in time we obtain an upper limit on the semi major axis and eccentricity right after the second collapse, $a_1$ and $e_1$. This, in turn, constrains the separation before the collapse, $a_0$, to $\min[a(1-e),a_1(1-e_1)]<a_0<a_1(1+e_1)\approx9.5\times10^{10}$\,cm.}}.

\citet[][hereafter BP16]{BP(2016)} used the observed orbital parameters of the Galactic DNS population  to constrain the distributions of mass ejection and kick velocities associated with the  formation of the second neutron star in the binary.
{While the smallness of the sample and unknown selection effects don't allow an accurate estimate of these distributions a clear picture emerges,} there are two
distinct types of neutron star formation. The majority of the systems, about two thirds, involve a minimal mass ejection ($\Delta M\lesssim 0.5M_\odot$) and 
low kick velocities ($v_{k}\lesssim 30$~km\,s$^{-1}$). The double pulsar system, 
PSR J0737-3039,  is a  prime candidate of this kind of collapse with $\Delta M =0.1-0.2M_\odot$ and $v_{k}=3-30 \mbox{\,km s}^{-1}$ \citep{Piran(2005),Dall'Osso2014}.
Such a  population of collapses with low mass ejection and kicks has been suggested on both observational \citep{Piran(2005),Wang2006,Wong2010,Dall'Osso2014} 
and  {theoretical \citep{DewiPols2003,Ivanova2003,VossTauris2003,Tauris2015} grounds.
{Subsequent  addition of a low mass ejection }channel of neutron star formation (via electron capture SNe) to population synthesis models 
\citep{Belczynski2008} improved the fit of the models to the observed DNS population.}
A large fraction of DNSs, born via the same mechanism, remain bound to their dwarf hosts. 
{On a related topic, \cite{RamirezRuiz2015ApJ} have argued that many DNSs are expected to remain confined and merge within globular clusters, which have comparable escape velocities to UFDs.} 

To explore these ideas we begin, in  \S \ref{sec:kicks} with a simulation of the typical  velocities of DNS systems using
the implied distributions of mass ejection and kicks from BP16.
We then address the delay times between formation and merger in \S \ref{rapid}. We show that a significant fraction of DNS systems will remain confined in UFDs and merge rapidly, demonstrating the viability of DNS mergers as sources of $r$-process material in UFDs.
In \S \ref{sec:impGalaxy} we consider the implications of these findings to the related problem of 
observation of heavy $r$-process material in some very low metallicity stars in the Galaxy. We summarize our results and the case for DNS mergers as the source of heavy $r$-process nucleosynthesis in \S \ref{sec:summary}.

\section{Confinement - Natal kicks of Double Neutron Stars}
\label{sec:kicks}
Consider the second collapse leading to the formation of the younger neutron star in a neutron star binary.
Assuming an initially circular orbit, the change in the centre of mass velocity , $\Delta \vec{v}_{CM}$, due to the second collapse is given by:
\begin{equation}
\label{eq:vcm}
 \Delta \vec{v}_{CM}=\frac{M_c}{M_p+M_c}\vec{v}_k+\frac{\Delta M}{M_c+M_p}\frac{M_p}{M_p+M_0}\vec{v}_{\rm kep} 
\end{equation}
where $ M_{p}$ and $M_{c}$ are the masses of the pulsar and the (collapsing)  companion today, $M_0$ is the companion mass right before the collapse, $v_{\rm kep}=\sqrt{G(M_0+M_p)/a_0}$ is the initial Keplerian velocity
and $\vec{v}_{k}$ is the kick velocity imparted on the companion by the ejected mass.
Naturally, if both the ejected mass, $\Delta M$ and the kick, $v_{k}$ are sufficiently small, the change in the CM velocity is also small.

{An inevitable mass loss, of order $\Delta M \geq \Delta M_{\nu}=0.1-0.15M_{\odot}$ \citep{LattimerPrakash2001}, arises due to emission of neutrinos. Plugging these values into
Eq. \ref{eq:vcm} with an initial separation $a=10^{11}$\,cm, final neutron star masses of $1.3M_{\odot}$
and no asymmetric kick, we find: $v_{CM,\nu}=11-16.5 \mbox{\,km s}^{-1}$. 
This is an approximate minimal value for the change in the CM velocity due to the collapse. A kick with a similar magnitude in approximately the opposite direction (or increasing $a$, although see \S \ref{sec:Introduction}) reduces somewhat this value.}

BP16 considered the orbital parameters and masses of the Galactic DNS population and constrained the two distributions of mass ejection and kick velocities directly from
observations (with no a-priori assumptions regarding evolutionary models and/or types of collapse involved). They've shown that there is strong evidence for two
distinct types of stellar collapses: The majority of systems have small eccentricities (we refer to these as ``small $e$ systems") implying a small  mass ejection ($\Delta M\lesssim0.5M_{\odot}$) and
a low kick velocity ($v_{k}\lesssim30$\,km\,s$^{-1}$).
Only a minority of the systems (``large $e$ systems") have been formed via the standard SN scenario involving a larger mass ejection
of $\lesssim2.2M_{\odot}$ and kick velocities of up to $400$\,km\,s$^{-1}$.

To calculate the fraction of systems with a given $\Delta \vec{v}_{CM}$ we performed a Monte Carlo simulation {using the best fit  distributions of collapse parameters found in BP16.}
{As mentioned earlier, }given the rather small number of observed DNS systems in the Galaxy and possible observational biases (see  BP16), it is {impossible to determine }the exact shapes of these distributions. The following calculations should thus be treated as an order of magnitude estimates. {Note, however, that  BP16 considered several functional shapes for these distributions and the results depend only weakly on the exact functional shape chosen.}

We assume that $60\%$ of the systems are ``small $e$ systems" with a log-normal mass ejection distribution {(above the minimum neutrino mass loss, which we take here to be $0.1M_{\odot}$)}
peaking at $\Delta M_0^{s}=0.05M_{\odot}$ and a log-normal kick distribution peaking at $v_{k,0}^{s}=5 $\,km\,s$^{-1}$ while $40\%$ of the systems are ``large $e$ systems" with a log-normal mass ejection distribution
peaking at $\Delta M_0^{l}=M_{\odot}$ and a log-normal kick distribution peaking at $v_{k,0}^{l}=158 $\,km\,s$^{-1}$. In all of these
distributions the standard deviations are half the magnitude of the corresponding peaks.
The direction of the kick velocities were chosen randomly. In addition we take $M_p=M_c=1.3M_{\odot}$, and initially circular orbits with constant separations.

\begin{figure*}
\centering
\includegraphics[scale=0.4]{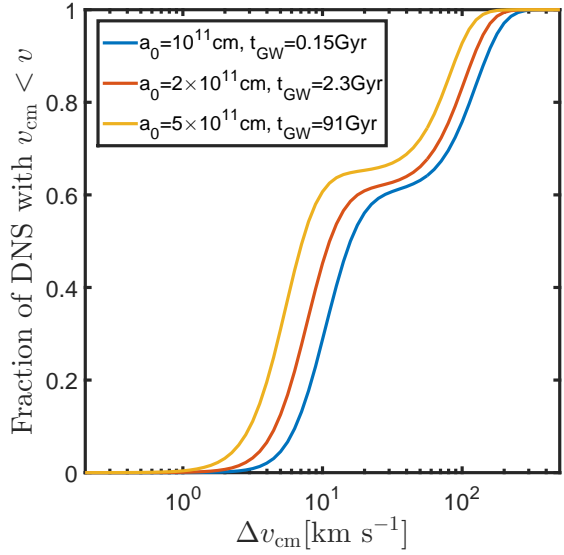}
\caption
{\small The fraction of DNS systems with $\Delta v_{\rm cm}<v$ as a function of $v$ for DNS distributions with different initial separations. $60\%$ of the systems are drawn from the ``small $e$ group" with a log-normal mass ejection distribution
{peaking at $\Delta M_0^{s}=0.05M_{\odot}$ above the minimum, $0.1M_{\odot}$ neutrino mass loss,} and a log-normal kick distribution peaking at $v_{k,0}^{s}=5 $\,km\,s$^{-1}$ while $40\%$ of the systems are drawn from the ``large $e$ group" with a log-normal mass ejection distribution
peaking at $\Delta M_0^{l}=M_{\odot}$ and a log-normal kick distribution peaking at $v_{k,0}^{l}=158 $\,km\,s$^{-1}$ (BP16). Due to the typical small kicks and mass ejection predicted by this model,
a significant fraction of systems (between $55-65\%$) have $v<15 $\,km\,s$^{-1}$. 
Such systems won't be ejected even from a UFD.}
\label{fig:vcmdist}
\end{figure*}

Fig. \ref{fig:vcmdist} depicts the fraction of DNS systems with $\Delta v_{CM}<v$ as a function of $v$ for different initial separations.
The initial Keplerian velocity decreases as the initial separation increases, resulting in a larger fraction of systems with small CM velocities for  larger separations. The shape of each curve is composed of two sharp rises, the first corresponding to the ``small $e$ systems" with weak kicks and the second to the ``large $e$ systems" with strong kicks. For initial separations in the range $a_0=10^{11}-5\times 10^{11}$\,cm,
a significant fraction of DNS systems (between $0.55-0.65$) have a CM velocity $<15 $\,km\,s$^{-1}$ 
(smaller than a typical escape velocity of UFDs).
This suggests that if these systems originated in such galaxies, they would, most likely, 
remain bound.

\section{Time until merger}
\label{rapid}
The time until merger of a DNS depends strongly on the initial separation and eccentricity.
Generally  an explosion occurring in one of the members of a binary, will cause the binary to become less bound.
Even if no asymmetric kick is imparted due to the explosion, the ejection of mass  causes the stars to be less gravitationally bound.
This can also be seen from the equation for
energy conservation with no kick:
\begin{equation}
 \frac{a}{a_0}=\left[2-\frac{M_0+M_p}{M_c+M_p}\right]^{-1}=\left[1-\frac{\Delta M}{M_c+M_p}\right]^{-1}>1\ .
\end{equation}
Nevertheless, if the  kick is oriented such that the relative velocity of the stars  decreases,  the system 
can become more bound after the explosion.
Since for a circular orbit, the kinetic energy is exactly minus that of the total orbital energy, the most energy that can be extracted is by reducing the kinetic to zero, thus
doubling the (negative) energy and leaving the binary with half of the initial semi-major axis and a very eccentric ($e\rightarrow 1$) orbit. 
For this to occur, the kick has to be oriented appropriately and its magnitude should be comparable to the Keplerian velocity.
A DNS system with a kick of this type would merge on a much shorter time-scale than implied by the original separation.

\begin{figure*}
\centering
\includegraphics[scale=0.4]{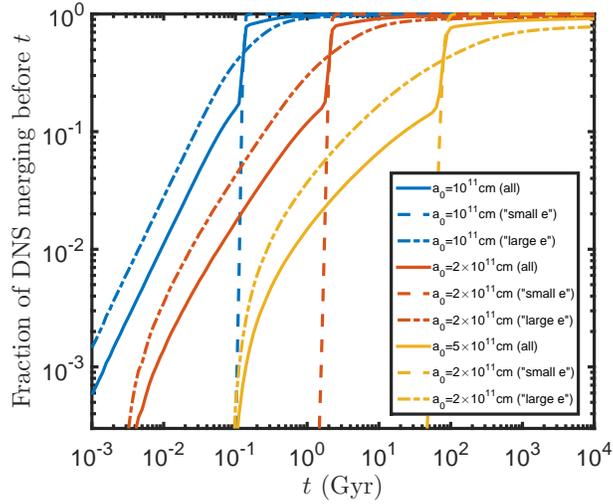}
\caption
{\small The fraction of DNS systems merging before time $t$ for the same initial separations.
The time for merger with no kick is $t_{GW}=0.15$\,Gyr for $a_0=10^{11}$cm ($t_{GW}=2.3$\,Gyr for $a_0=2\times10^{11}$\,cm).
The distribution for ``large $e$ group" (``small $e$ group") is plotted in dot-dashed (dashed) lines.
Solid lines depict the distribution for the general population, assuming that $0.6$ of the systems are drawn from the ``small $e$ group" and $0.4$ of the systems are drawn from the ``large $e$ group".
Due to the typical large kicks and mass ejection predicted for the ``large $e$ group", a significant fraction of these systems can merge much faster than implied by their original orbits.}
\label{fig:probtmerger}
\end{figure*}

We carried out a Monte Carlo simulation as described in \S \ref{sec:kicks}.
For each realization, we calculate the conditions after the second collapse and use the separation and eccentricity to find the time until merger due to GW emission.
The fraction of DNS systems merging by a time $t$ for both the ``small $e$" and ``large $e$" groups 
is shown in Fig. \ref{fig:probtmerger}.
For $a_0=10^{11}$\,cm ($a_0=2\times10^{11}$\,cm), corresponding to a merger time of $t_{GW}=0.15$\,Gyr ($t_{GW}=2.3$\,Gyr), 
$10\%$ of ``large e" systems
(corresponding to $4\%$ of the entire DNS population) merge in less than $25$\,Myr ($0.8$\,Gyr)
{\citep[see][]{Belczynski(2002),Belczynski2007,O'Shaughnessy2008ApJ}}. 
This population  has important effects on the typical delay between SNe and DNS mergers leading to 
a faster rise in the abundance of heavy r-process elements as a compared to the iron abundance \citep{shen2015ApJ}.

\begin{figure*}
\centering
\includegraphics[scale=0.4]{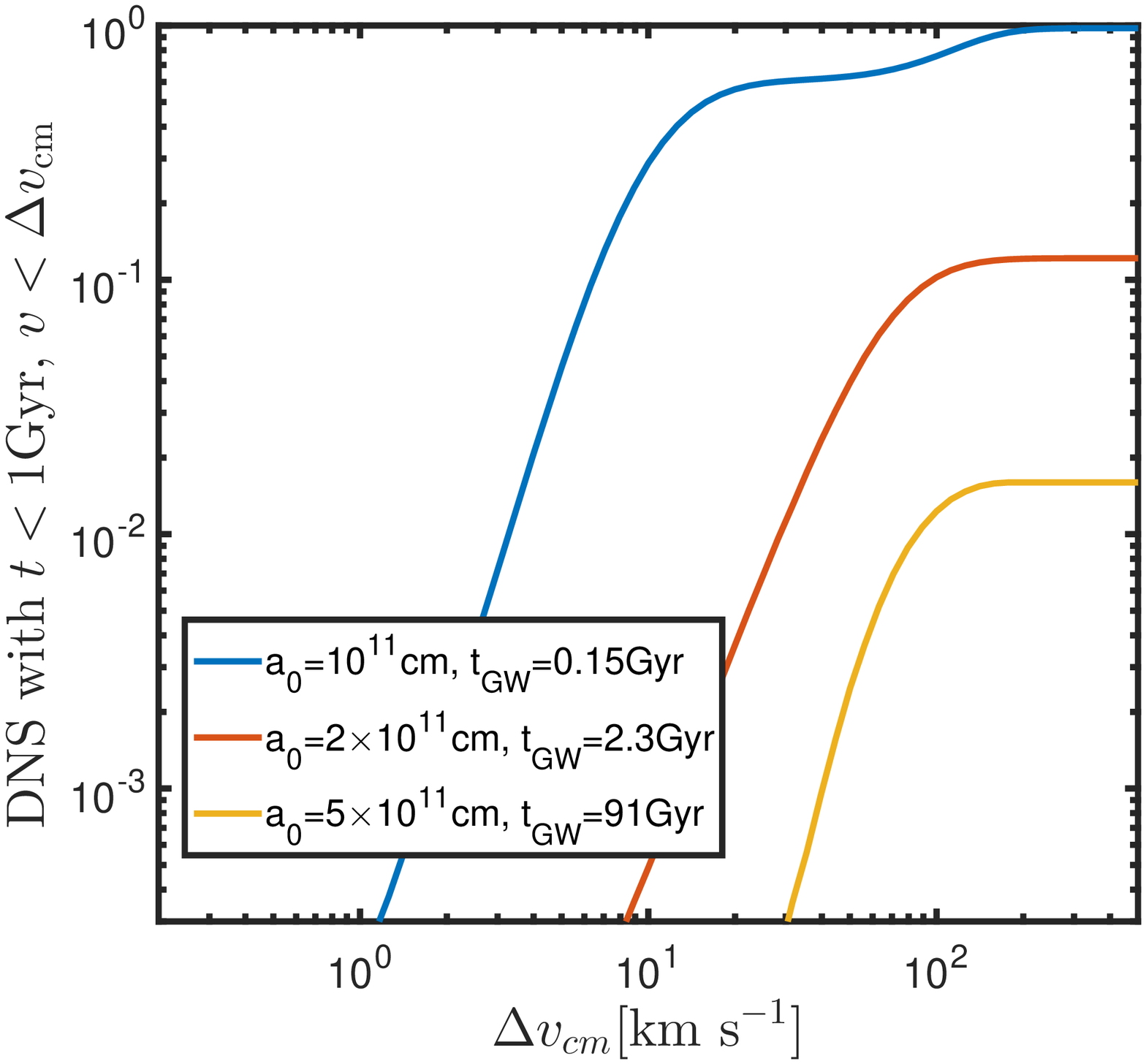}
\caption
{\small The fraction of DNS systems merging by a given time $t<1$\,Gyr and with $\Delta v_{\rm cm}<v$.
The time for merger with no kick is $t_{GW}=0.15$\,Gyr for $a_0=10^{11}$\,cm ($t_{GW}=2.3$\,Gyr for $a_0=2\times10^{11}$\,cm).
Since systems with confining kicks preferentially correspond to larger changes in the CM velocity, the main requirement
for obtaining such systems is for the initial separation to correspond to a merger at $t<1$\,Gyr.}
\label{fig:probtandv}
\end{figure*}

So far we have addressed the questions of merger time and CM velocities separately. However, systems with confining kicks are preferentially those that undergo larger changes in the CM velocity. The decrease in the gravitational orbital energy is compensated by an increase in the kinetic energy of the CM. Thus, although important for larger galaxies, confining kicks are not relevant for UFDs. For the latter, it is important to consider the combined probability of having a rapid merger with a small CM velocity. 
Fig. \ref{fig:probtandv} describes the fraction of DNS systems merging before 1\,Gyr with $\Delta v_{\rm cm}<v$ as a function of $v$.
We  consider an initially circular orbit with a separation $a_0=10^{11}$\,cm ($a_0=2\times10^{11}$\,cm). 
For $a_0=10^{11}$\,cm, $55\%$ of the systems merge by 1\,Gyr and have $\Delta v_{\rm cm}<15 $\,km\,s$^{-1}$ (comparable to the lowest escape velocities of UFDs). However, this fraction decreases significantly, down to $0.17\%$, for $a_0=2\times10^{11}$\,cm. 
As expected the distribution is dominated by the ``small $e$" group of systems which merge on a time-scale that is dictated by their initial separation.

The existence of rapid mergers as well as evidence for the typical initial separations used here is supported by observations of known DNS systems in our Galaxy (see \S \ref{sec:impGalaxy} for details).
Another line of evidence for the existence of rapid mergers arises from observations of sGRB hosts.
Since sGRBs are most likely associated with DNS mergers \citep{eichler1989Nature}, one may expect that in the absence of rapid mergers, sGRBs would be predominantly observed in the more massive elliptical galaxies. In fact, a significant fraction of sGRBs are observed in star forming galaxies \citep{Fong2013ApJ}.
The prevalence of sGRBs in star forming galaxies is naturally expected if a significant population of DNS systems merge rapidly.

\section{Implications for Galactic nucleosynthesis} 
\label{sec:impGalaxy}
Our main aim in this paper is to explore the possibility of a DNS mergers taking place in UFDs. 
Still, this work and in particular the parts corresponding to rapid mergers has also implications to nucleosynthesis in the Milky Way. 
Abundances of $r$-process elements have been observed in halo and disk
stars covering a metallicity range ${\rm[Fe/H]}\approx-3.1-0.5$
\citep{woolf95,sneden00,Cayrel01}.  
This requires a significant fraction of $r$-process nucleosynthesis to take place within a few hundred Myr or less. 
This has been used as an argument against DNS mergers as the origin of heavy $r$-process material in the Galaxy \citep{argast2004A&A}. The argument is roughly the following: DNS mergers will take place a few Gyr after star formation at which stage all stars would already be ``contaminated" by a significant amount of Iron. Furthermore, DNS mergers require SNe  
and those will produce iron before the $r$-process material is produced.

{Most Galactic} DNS systems were born in the past when the Galactic SFR was larger \citep[e.g.][]{majewski1993ARA}. {However,} all the ``rapid mergers" that formed at that time have already merged. Hence the total fraction of rapid mergers is larger than what is currently observed in the Galactic binary pulsar population. This result is consistent with the global fast decay in the rate of sGRBs in the Universe \citep{wanderman2015MNRAS}. It is also consistent with 
the observation that the current rate of deposition of radioactive $^{244}$Pu on Earth is much lower than what it was 4.6~Gyr 
ago when the solar system was formed \citep{Wallner2015NatCo,hotokezaka2015NaturePhys}.

We have shown here that DNS mergers can take place rapidly enough and enrich 
at least some parts of our Galaxy on very short time scales. Specifically, $10\%$
of the large $e$ systems with $a_0=10^{11}$\,cm have merger times less than $25$~Myr.
They play important roles to account for a fast rise in the abundances of heavy $r$-process elements
of metal poor stars in the Galaxy.
The large fluctuations in the Eu/Fe ratio in metal poor stars, supports a rare enrichment process like DNS 
merger~\citep[e.g.][]{Cescutti2014,tsujimoto2014A&A,wehmeyer2015MNRAS}. 
Regarding the problem of ``contamination" of the ISM by iron from SNe that took place prior to the formation of the DNS system, 
we note that in a significant fraction of the cases, the DNS systems are produced with very little mass ejection. 
In other cases, the proper motion of the DNS relative to their birth rest frame suggests that even in a few hundred Myr they could move sufficiently within the Galaxy to avoid the ISM that has been contaminated by their ``progenitor" SNe. Such an effect should be taken into account in chemical evolution studies. An open question related to this involves the not yet understood topic of possible large scale turbulent mixing in the galaxy.

\section{Summary}
\label{sec:summary}
We have examined here DNS mergers as sources of heavy $r$-process material in dwarf galaxies.
We have shown  that both arguments raised by \cite{Bramante(2016)}, that DNS systems may be ejected from their galaxies due to strong kicks received at formation, and that DNS would not be able to merge rapidly enough before the star formation in these galaxies has stopped, are naturally overcome. First, due to a significant population of DNS receiving weak kicks at birth (BP16), a large fraction of DNS systems would have CM velocities $<15 $\,km\,s$^{-1}$ (comparable to the lowest escape velocities from UFDs).
Second, given limits on the separation of the double pulsar system before its second collapse, a significant fraction of systems
are expected to both remain confined in UFD galaxies and merge within less than a Gyr.

Moving from UFDs to the Galaxy, where small velocities are not  required to remain confined, the fraction of rapid mergers becomes even larger due to the contribution of ``confining kicks". Limits on the time between the second collapse until eventual merger, of two out of the ten observed DNS systems in the galaxy, imply that many DNS systems are expected to merge within less than a few hundred Myr.
This is also supported by observations of a rapid decay in the  sGRB rate following the peak of star formation in the universe.
This population of rapid mergers is composed mainly of systems that were formed with small amounts of mass ejection $\approx 0.1M_{\odot}$, (BP16). Moreover,  the mergers take place  far away from the place of the progenitor SNe. This implies that DNS mergers can naturally account for the observations of significant amounts of $r$-process material  in some low metallicities halo stars in the Galaxy. This invalidates the argument used against DNS mergers as the origin of $r$-process nucleosynthesis in our Galaxy \citep{argast2004A&A}.
In fact, DNS mergers are more easily compatible with the large fluctuations in the Eu/Fe ratio in metal poor stars which are more easily accounted for by rare events. In a companion paper \cite{Beniamini2016} we provide further quantitative evidence that $r$-process material is produced in rare events in both our own Galaxy and UFDs. 

We thank Nicholas Stone {and Todd Thompson}
for useful discussions and comments. 
This work was supported in part by an Israel Space Agency (SELA) grant, the Templeton foundation and the I-Core center for excellence ``Origins" of the ISF.

\hyphenation{Post-Script Sprin-ger}
\hyphenation{Post-Script Sprin-ger}

\end{document}